# Neural Network Training for OSNR Estimation From Prototype to Product

Andrew D. Shiner[1*], Mohammad E. Mousa-Pasandi[1], Meng Qiu[1], Michael A. Reimer[1], Eui Young Park[1], Michael Hubbard[1], Qunbi Zhuge[1,2], Francisco J. Vaquero Caballero[3,1], Maurice O'Sullivan[1]

[1] *Ciena Corporation, Ottawa, Ontario Canada*
[2] *Shanghai Jiao Tong University, Shanghai, China*
[3] *Cambridge University, Cambridge, U.K.*
[*] *ashiner@ciena.com*

**Abstract:** A method for in-service OSNR measurement with a coherent transceiver is presented and experimentally verified. A neural network is employed to identify and remove the nonlinear noise contribution to the estimated OSNR. © 2020 The Author(s)

## 1. Introduction:

There is increased demand for modem-based performance measures to assess the operation and capability of networks. A longstanding measure of performance is the link delivered optical signal to noise ratio (OSNR), depicted in Fig. 1A, where for OSNR the "signal" power contains all in-band components except for ASE and the noise is the integrated ASE power spectral density (PSD) over an agreed bandwidth (typically 0.1 nm). The OSNR is usually measured with an optical spectrum analyzer (OSA) where the ASE PSD is estimated from the spectrum beside the modulated signal. This measurement becomes challenging in wavelength division multiplexed (WDM) systems where closely spaced channels make accurate estimation of the out of band ASE noise power difficult. In high spectral efficiency near Nyquist channel WDM transmission there is little if any ASE-only floor region to measure.

An OSNR estimator has been developed which is compatible with in-service use in a commercial coherent transceiver. The OSNR is calculated using the effective signal to noise ratio (ESNR) as well as the external SNR ($SNR_{ext}$). ESNR is the effective additive white Gaussian noise (AWGN) SNR obtained by inverting the measured pre-FEC BER. $SNR_{ext}$ is calculated from the ESNR by removing the modem noise contribution [1], and represents the modem estimate of the SNR contributions from all noise sources external to the modem. We note that external SNR is not strictly modem independent as the noise developed in the line depends on modem parameters such as constellation design [2], amount of electronic pre-dispersion [3], use of nonlinear pre-distortion etc. $SNR_{ext}$ is calculated, (linear units throughout), as:

$$\frac{1}{SNR_{ext}} = \frac{1}{EC \cdot ESNR} - \frac{1}{SNR_{imp}}, \qquad (1)$$

where signal to implementation noise ($SNR_{imp}$) and eye-closure (EC) are modem calibration factors. To estimate the OSNR we first estimate the $SNR_{ASE}$ which is related to OSNR by the ratio of the symbol rate ($B_e$) to the OSNR reference bandwidth ($B_o$), as $OSNR = SNR_{ASE} \left(\frac{B_e}{B_o}\right)$. The $SNR_{ASE}$ is calculated from the $SNR_{ext}$ by removing the total nonlinear noise SNR ($SNR_{NL}$) as:

$$\frac{1}{SNR_{ASE}} = \frac{1}{SNR_{ext}} - \frac{1}{SNR_{NL}}. \qquad (2)$$

The relationship between the external and nonlinear SNRs has a complicated dependence on fiber topology and power profile. A neural network (NN), described in the next section, can be used to estimate $SNR_{NL}$ from correlation properties of the received symbols.

## 2. Nonlinear Noise Estimation:

A modem can report received symbols from its soft decoder. The noise on each received symbol is estimated from the difference between that symbol and the estimate of the corresponding transmit symbol. This noise field contains contributions from ASE, fiber nonlinearity and decision errors among other sources. In order to estimate the contribution from $SNR_{NL}$ we exploit the temporal and polarization correlations that are imposed on the nonlinear noise field by the Kerr nonlinear processes. We consider nonlinear noise contributions from intra-channel self-phase modulation (SPM) as well as inter-channel cross-phase modulation (XPM) and cross-polarization modulation (XPolM). The SPM contribution is estimated from C-coefficients [3,4], and the XPM and XPolM contributions from correlation doublets [5] between co-polarized and cross-polarized symbol pairs respectively. A NN comprising 35 inputs, 2 hidden layers with 5 nodes per layer was trained to discover the relationship between the $SNR_{NL}$ and the C-coefficients, doublets and other parameters measured by the modem such as ESNR, link length and cumulative dispersion.

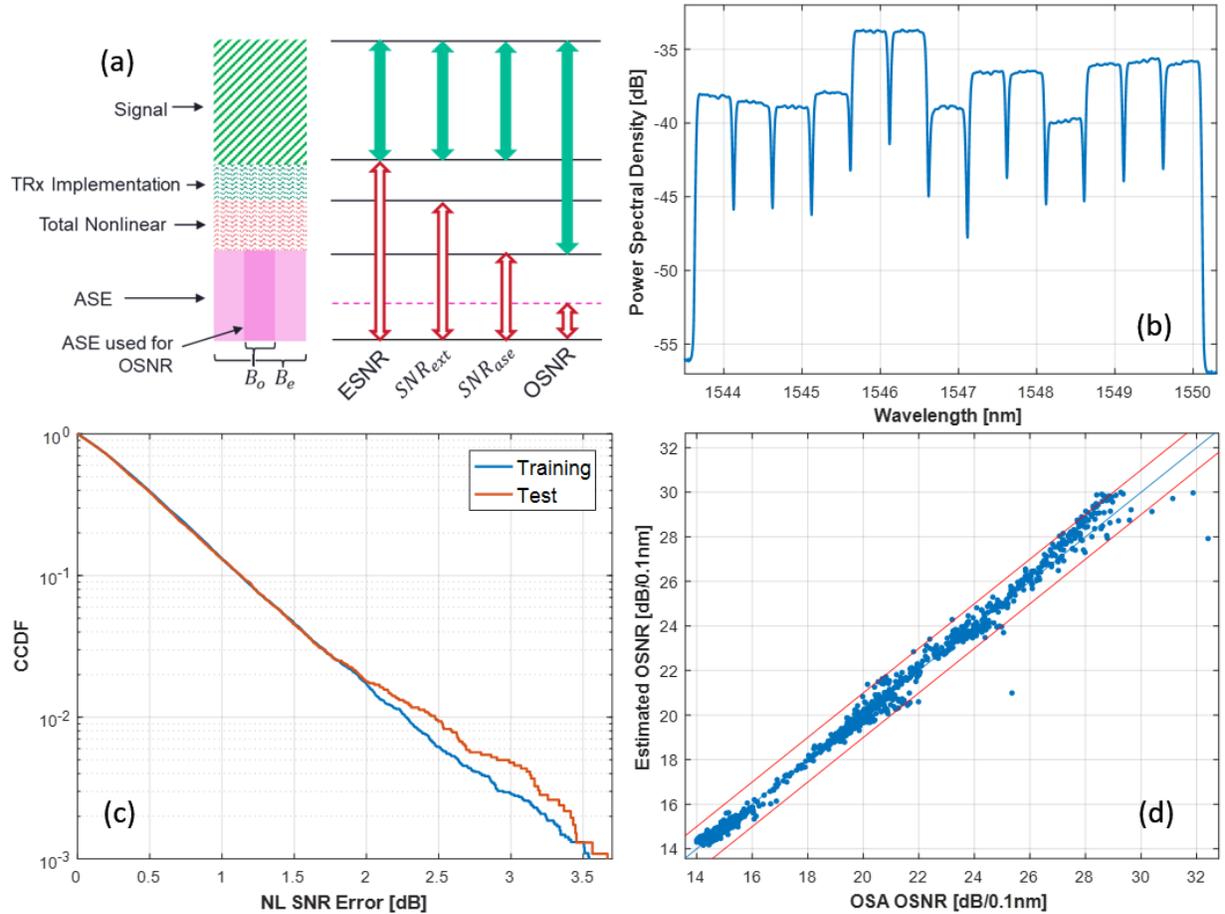

*Figure 1(a) Signal (green arrows, top) and noise sources (red arrows, bottom) accounted for in different SNR metrics including OSNR. Note that the OSNR uses different integration bandwidths for signal and noise. (b) OSA spectrum of 13-channel steps model following propagation. (c) NN training results for 56-200G propagation showing the complementary cumulative distributions (CCDFs) for the magnitude of the error between the estimated $SNR_{NL}$ from the NN vs direct calculation from the simulated nonlinear field. (d) WLAi estimated OSNR vs OSA measurements combining results for back-to-back and propagated cases over NDSF and TWc fiber with single channel, equal power WDM and 13-channel steps model conditions.*

A large training/test ensemble of diverse channel power and fiber configurations was simulated and the so-trained NN was validated with measured input vectors and OSNRs. Simulations were based on split-step Fourier (SSF) solutions of the Manakov equation. A hardware respecting transmitter model generated baseband waveforms which were multiplexed onto the transmit spectra with other channels as needed. Simulations consisted of single channel as well as N-channel (N ≈ 13) cases using constant launch power for all channels or powers drawn from the N-channel steps model as follows. In the steps model N channels are divided into $M \sim \mathcal{U}\{1, N\}$ adjacent groups, where $\mathcal{U}\{a, b\}$ is the discrete uniform distribution between a and b. The number of channels per group is chosen from $\mathcal{U}\{1, N\}$ with the added constraint that the total number of channels for all groups is N. The launch power for each group is normally distributed, $\mathcal{N}(\mu, \sigma)$, with mean $\mu$ given by the optimal launch power from the GN model [6], less a constant offset (typically 2 dB). The standard deviation $\sigma \sim 2$ dB was chosen to generate variability in power among groups. Repeated application of the steps model generates training cases with nearly flat power profiles, individual channels that are launched with significantly higher powers than their neighbors and everything in between. An example of a 13-channel steps model power profile that was measured in a lab with an OSA is shown in Fig 1B. For mixed fiber type simulations, the WDM power spectrum is scaled at each amplifier site by the ratio of the optimum launch powers for the following and preceding fiber spans. Random fiber realizations were generated from a fiber model designed to mimic known inter-exchange carrier routes. The model consisted of 7 fiber types (ELEAF, LEAF, NDSF, TERALIGHT, TWc, TWP, TWRS) with associated probability distribution functions (PDFs) that capture the distribution of span lengths by fiber type. A transition matrix selects the fiber type for the next span based on the current span with appropriate probabilities.

Typical WDM simulations consisted of 13 independently modulated carriers each encoding ~$2^{16}$ symbols per carrier. Following propagation each carrier was brick-wall filtered, frequency shifted to baseband and down sampled. At this stage the waveform for each channel was patterned ~20 times and passed through a linear channel emulator model consisting of a finite number of emulator segments (typically 50) where each segment applies a random polarization rotation, as well as random amounts of polarization mode dispersion (PMD), polarization dependent loss (PDL) and ASE. The PMD and PDL values were drawn from Maxwellian distributions where PMD is parameterized by the mean differential group delay (DGD) with typical values of (DGD = 25 ps, PDL = 2 dB) for 200G, (DGD = 16 ps, PDL = 1.3 dB) for 300G and (DGD = 5 ps, PDL = 0.67 dB) for 400G transmission. The SNR from ASE was chosen for a given propagated wave, and the noise was uniformly distributed between the emulator segments. In most cases the SNR from ASE was chosen uniformly between the link delivered SNR and the SNR at the FEC threshold. The link delivered SNR was estimated for each link condition using an amplifier model with nominal noise figure and gain. The receive DSP was implemented in floating point and was designed to mimic the behavior of the major functional blocks that are presented in a commercial coherent transceiver which include electronic dispersion compensation, polarization, carrier phase and clock recovery and symbol decoding.

### 3. NN Training Results:

An example of results from the NN training set that was used for 56 GS/s, 200 Gb/s (56-200G) transmission is shown in Fig. 1C. The training/test set consisted of 23900 propagated channels from which 5274 were reserved for testing and the rest were used for training. In simulation the NN estimation accuracy for $SNR_{NL}$ using the test set data had a systematic offset of +0.07 dB with a standard deviation of 0.63 dB.

### 4. Verification:

The OSNR estimator was implemented in Ciena's WLAi coherent modem which solved Eq. (2) using estimates for $SNR_{NL}$ from a firmware implementation of the NN described above. The test channel propagated alone or was surrounded by 6 channels on either side from a WLAi bulk modulator. Optical powers were swept for single channel and WDM equal-power configurations, or were drawn from the 13-channel steps model. The fiber plant consisted of up to 30 spans of non-dispersion shifted fiber (NDSF), or up to 10 spans of True Wave Classic (TWc) fiber, each with a nominal span length of 80km. The propagated test channel was optically noise loaded at the receiver prior to decoding with a WLAi receiver which reported its estimate of the OSNR. The estimated OSNR was compared with the OSNR measured with an OSA at the input to the receiver. In most cases we measured at the link delivered OSNR as well as with noise loading to a constant BER of 3.2%. The firmware implementation includes a step which detects unrealistic link conditions where the $SNR_{NL}$ is outside of the range which would be expected based on the $SNR_{ext}$ and flags the OSNR estimates as invalid. Note that in some cases the firmware processing steps were implemented offline with symbols returned from the soft decoder. We verified equivalency between the firmware and offline OSNR calculations. Results shown in Fig. 1D combine WLAi 56-200G, 56-300G and 56-400G transmission modes over a range of fiber type, distance and channel configurations. The combined accuracy for each transmission mode is expressed as a (systematic shift $\pm$ σ) of (0.04 $\pm$ 0.25) dB, (-0.03 $\pm$ 0.29) dB and (-0.12 $\pm$ 0.48) dB for WLAi 56-200G, 56-300G and 56-400G modes respectively.

### 5. Conclusion:

An OSNR estimator based on a neural network has been implemented in Ciena's WLAi coherent transceiver. This paper described the procedure for simulation-based training and presents experimental verification results.